\documentstyle[prd,aps]{revtex}
\input psfig
\begin{document}
\draft
\twocolumn[\hsize\textwidth\columnwidth\hsize\csname@twocolumnfalse%
\endcsname
\title{High Energy Astrophysical Neutrinos: the Upper Bound is Robust}
\author{John Bahcall}
\address{Institute for Advanced Study, Princeton, NJ 
08540\thanks{jnb@sns.ias.edu}}
\author{Eli Waxman}
\address{Department of Condensed-Matter Physics, Weizmann Institute of 
Science, Rehovot 76100,
Israel\thanks{E-mail:waxman@wicc.weizmann.ac.il}}
\date{\today}
\maketitle

\begin{abstract}
We elucidate the physical basis for the upper bound on high energy
neutrino fluxes implied by the observed cosmic ray flux.  We stress
that the bound is valid for neutrinos produced either by $p,\gamma$
reactions or by $p-p(n)$ reactions in sources which are optically thin
for high energy protons to photo-meson and nucleon-meson interactions.
We show that the upper bound is robust and conservative.  The
Waxman-Bahcall bound overestimates the most likely neutrino flux by a
factor $\sim 5/\tau$, for small optical depths $\tau$. The upper limit
cannot be plausibly evaded by invoking magnetic fields, optically
thick AGNs, or large hidden fluxes of extragalactic protons.  We
describe the implications of the bound for future experiments
including the AMANDA, ANTARES, Auger, ICECUBE, NESTOR, and 
OWL/AIRWATCH detectors.
\end{abstract}

\vspace*{.1in}
\pacs{PACS numbers: 95.85.Ry, 98.70.Rz, 98.70.Sa, 14.60.Pq}

]


\section{Introduction}
\label{sec:introduction}

The observed cosmic ray flux at high energies implies an upper bound
on the high-energy neutrino flux produced in astronomical sources that
are, like gamma ray bursts and the observed jets of active galactic
nuclei, optically thin to photo-meson interactions on
protons~\cite{WnB99}.  
The upper bound also applies to 
$p-p$ or $p-n$ collisions that create  neutrinos via pion production in
sources that are optically thin to nucleon-meson interactions.
For simplicity, we shall refer in the following only to $p-p$
collisions when we really mean either $p-p$ or $p-n$ collisions.

The high energy protons that produce neutrinos by $p-\gamma$ or by
$p-p$ interactions contribute to the observed cosmic ray flux after
they leave the site where the neutrinos are created.  Since they have
maintained a high energy in spite of the possibility of interactions
with the cosmic microwave background, protons that are detected at
earth with energies greater than $ 10^{18}$~eV must originate at
redshifts $z < 1$.  Therefore, the observed flux of high-energy cosmic
rays determines the rate at which particles of those energies are
being created in the relatively local universe.  One can obtain an
upper limit to the cosmic ray production rate in the whole universe by
assuming that the rate increases with redshift like the fastest known
population of astronomical sources, quasi-stellar sources (QSOs).
Knowing an upper limit to the universal proton production rate, one
can readily compute, by standard particle physics techniques, an upper
limit to the rate of production of neutrinos by the same protons.

The conservative upper bound derived in this way has come to be known
in the literature as the Waxman-Bahcall upper bound~\cite{WnB99}.  

The upper bound exceeds what one can reasonably expect to measure. As
we shall see in Sec.~\ref{sec:crandWB}, the
limit is derived numerically by assuming that all of the energy of the
high energy protons produced in the astronomical sources is
transferred to pions by photo-pion or nucleon-proton interactions.
Each time that a $\pi^+$ is produced by photo-pion interactions it
receives only about $20$\% of the initial proton
energy~\cite{WnB99}. For $p-p$ collisions at very high energies, the
explicit calculation of the fraction of energy that is transferred to
pions has not yet been done but is likely to be even less than for
$p-p$ interactions.  Thus, at the very least, the Waxman-Bahcall upper
bound is conservative by a factor of five, and 
the most likely neutirno flux is about $5/\tau$
smaller than the Waxman-Bahcall limit for small optical depths.

The Waxman-Bahcall bound 
is consistent with our prediction of the expected flux
of high energy neutrinos from gamma-ray bursts (GRBs)\cite{WnB99,WnB97}.
Naturally,
the flux expected from GRBs is less than the maximum allowed by the bound.
For the GRB calculation, we estimated  that $\sim 20$\% of the total proton 
energy is transferred by photo-mesonic interactions to pions 
(proton-nucleon interactions are less efficient~\cite{BM2000}).  
As described in the previous paragraph, we 
assumed, in deriving our upper bound, 
that $100$\% of the proton energy is transferred to pions.

The upper bound we set from the observed cosmic ray flux is 
two orders of magnitude lower than the intensity
predicted in some  previously published 
 models for the production of neutrinos in AGN jets, implying 
that a  km$^2$ 
neutrino detector would record  at most $1$  neutrino from AGN jets 
per year. 
The same argument also rules out 
models in which most or all of the gamma-ray background is produced by
photo-meson interactions in AGN jets.

Can the upper bound on the high energy neutrino flux be evaded?  This
is an important question since very large neutrino detectors are being
designed for installation in the ocean or a deep lake (see,
e.g., Refs.~\cite{antares,nestor,baikal}), under Antarctic ice (see, 
e.g., Ref.~\cite{amanda}), in space (see, e.g., Ref.~\cite{owl}), and
large area ground arrays (see, e.g., Refs.~\cite{auger,watson}).
The design characteristics for these neutrino detectors are being
determined in part by the best available theoretical models. It is
therefore necessary to examine carefully the justifications for
different predictions of high energy neutrino fluxes.  We do so in
this paper.

The goal of this paper is to clarify how the Waxman-Bahcall upper
bound follows from the measured cosmic ray flux and, in the process of
the clarification, to demonstrate that the upper bound is robust and
conservative.  We emphasize the  implications of the bound
for future neutrino and gamma-ray observations.

There are two special types of sources for which the Waxman-Bahcall
bound does not apply; these sources could in principle produce a
neutrino flux exceeding the Waxman-Bahcall limit. The first special
type of source is one in which neutrinos are produced by processes
other than photo-meson or proton-nucleon interactions; the second type
of special source is one for which the photo-meson (or proton-nucleon)
optical depth is high.  We begin by summarizing in
 Sec.~\ref{sec:exceed} various speculations regarding the existence
of such sources, for which we have no observational evidence to
date. This section is particularly relevant for future neutrino
observational programs. We then turn to a detailed discussion of the
derviation of the bound and of its validity and robustness.

In Sec.~\ref{sec:crandWB} and
Fig.~\ref{fig:CR}, we summarize the existing data on the cosmic energy
spectrum, including the limited information at very high energies that
is derived from the Fly's Eye~\cite{fly}, AGASA~\cite{agasa}, and
Yakutsk~\cite{yakutsk} experiments.  We show in this section
that cosmic-ray observations imply a
a robust upper limit on neutrino fluxes in the energy range of 
$\sim10^{16}$~eV to $\sim10^{20}$~eV.
What is expected at very high energies, $ > 10^{20}$ eV, and at
relatively low energies, $< 10^{16}$ eV?  We argue in
Sec.~\ref{sec:nogzk} that it is plausible to extrapolate the
Waxman-Bahcall bound to neutrino energies that correspond to protons
more energetic than the expected Greisen-Zatsepin-Kuzmin cutoff at $5
\times 10^{19}$ eV.  In Sec.~\ref{sec:lowenergy}, we summarize the
observational evidence for the conventional viewpoint that most cosmic
rays below $10^{18}$~eV are heavier particles of Galactic origin.
Thus, the proton flux that is to be used in deriving the upper limit to
the neutrino flux is much less than the total observed cosmic ray flux
at energies below $10^{18}$~eV. We discuss the important question of whether 
sources that are not included in the Waxman-Bahcall bound derivation
may exist in this energy range.

It has been common practice for some time to motivate proposed
observatories to search for high energy neutrinos by considering the
predictions of AGN models for neutrino production for photo-meson
interactions that also produce the gamma-ray background.  We show in
Sec.~\ref{sec:AGN} that the large set of previously published AGN
models for neutrino production exceed the Waxman-Bahcall bound by
typically two orders of magnitude and therefore are not appropriate
theoretical models upon which to base proposed observatories.  We
present estimates for the upper limit neutrino event rates photo-meson
and nucleon-meson interactions that may be observed in the AMANDA, ANTARES,
Auger, ICECUBE, NESTOR, and OWL detectors.  The previously
published AGN models based upon photo-meson interactions are also
inconsistent with gamma-ray observations that show rapid time
variability. We also discuss in this section the question of whether
some AGN's models may have a high photo-meson optical depth.

Can one increase the rate at which high energy protons are generated
beyond the value contemplated in the Waxman-Bahcall bound? If so, one could 
obviously raise the experimentally important 
bound on the neutrino flux.
The only way of exceeding the bound 
is by hypothesizing the existence of
high energy proton accelerators 
 which exist in environments that do not allow
 protons to escape.\footnote{We point out in
Ref.~\cite{WnB99} that one can imagine ``neutrino only'' 
sources that are  optically thick to proton photo-meson
interactions and from which  protons  cannot  escape.
Sources of this kind
could in principle contribute a flux in excess of the Waxman-Bahcall
bound, but there is, by construction, 
no observational evidence (from baryons or high energy 
photons) for their existence.}
Can one fine-tune the conditions in which  luminous astronomical sources are
imagined to exist so that  neutrinos and photons escape, 
but  protons do not leave the system?
Strong magnetic fields obviously provide a potential way of confining
protons, but not neutrons, 
and we discussed this possibility briefly in Ref.~\cite{WnB99}. 
In Sec.~\ref{sec:magnetic}, we consider some specific scenarios for avoiding 
the Waxman-Bahcall limit by invoking strong mangetic fields and show by 
example 
that magnetic field confinement is not a plausible way of aoviding the limit.

In Sec.~\ref{sec:nonresonant} we show that it is a good
approximation for GRB's to assume that the high energy neutrino flux
is dominated by photo-pion interactions that proceed through the $\Delta$
resonance. The contributions from non-resonant itneractions are shown to 
be small, comfirming our previous estimates \cite{WnB97,WnB99}, according 
to which $\sim20\%$ of the proton energy is converted in GRBs to pions.
We summarize our results in Sec.~\ref{sec:discussion}.

\section{Exceed, not violate, the WB bound}
\label{sec:exceed}

Does the Waxman-Bahcall bound imply that neutrino fluxes from all
conceivable astronomical sources must lie below this limit? Are all
possible neutrino sources that will be searched for with very large
area neutrino telescopes subject to this bound?

As we discussed in the Introduction, there are two (speculative) ways
to exceed the limit without violating the constraint imposed by the
observed cosmic ray background. The WB bound refers, see
Sec.~\ref{sec:crandWB}, only to sources that are optically thin to
proton photo-meson and proton-nucleon interactions and from which
protons can escape.  Thus, the first `way out' which was also discussed in
our original paper~\cite{WnB99}, is to assume  the existence of sources
which are optically thick to proton photo-meson or proton-nucleon
interactions. We referred in~\cite{WnB99} to such sources as ``neutrino
only factories'' or ``hidden core models'' since they are not
motivated by measurements of 
the cosmic-ray flux or by any  electromagnetic
observations. 

Examples of optically thick scenarios include neutrinos produced in
the cores of AGNs (rather than in the jets) by photo-meson
interactions~\cite{Stecker}, or via $p-p$ collisions in a collapsing
galactic nucleus~\cite{BD00} or in a cacooned black
hole~\cite{BG87}. The optimistic predictions of the AGN core
model~\cite{Stecker} have already been ruled out by the AMANDA
experiment \cite{amanda-bound} (see Fig.~\ref{fig:agnmodels}).
 
The second type of source that may exceed the WB bound is one 
in which neutrinos are produced by some mechanism other than 
photo-meson or proton-nucleon interactions. We list below some of the
hypothesized sources of this type. 

\begin{itemize}

\item Decays of supermassive, dark matter
particles~\cite{ellis1981,berezinsky97} might produce high energy
neutrinos without violating the cosmic ray bound.

\item Topological defects (see, e.g., Refs.
\cite{Vilenkin85,Hill85,Schramm92,BV99,Halzen2000,LS2000}) might
produce particles which decay, among other channels, to neutrinos with
only a small associated cosmic ray proton flux.

\item Superheavy relic neutrinos~\cite{Weiler99,Gelmini2000} which
interact with the cosmic neutrino background have been proposed as one
way of producing ultra-high energy cosmic rays.

\item Ultrahigh-energy photons at large redshifts~\cite{kusenko2000}
could in principle produce a flux of ultrahigh-energy neutrinos.

\end{itemize}

\section{Cosmic ray observations and the WB bound}
\label{sec:crandWB}

In this section, we summarize the observations and calculations that
lead directly to the upper bound on high energy neutrino fluxes.

\begin{figure}[!t]
\centerline{\psfig{figure=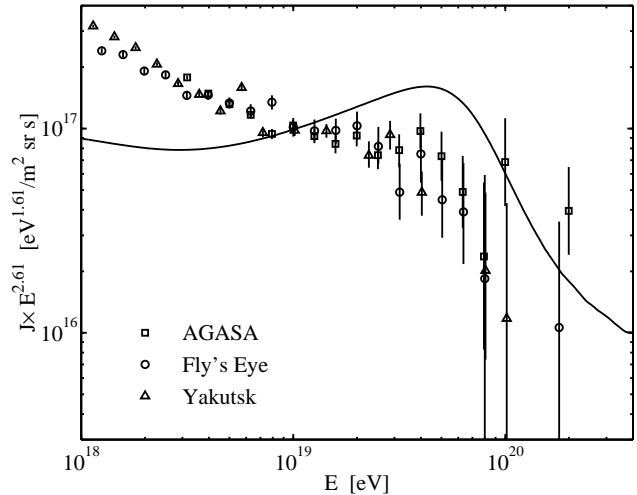,width=3.3in}}
\caption[]{The observed high energy cosmic ray flux. Measurements are
shown from the Fly's Eye~\cite{fly}, AGASA~\cite{agasa}, and
Yakutsk~\cite{yakutsk} detectors. The smooth curve, computed from
Eq.~(\ref{eq:ECR}), was used by Waxman and Bahcall~\cite{WnB99} to
compute the upper bound on high energy astrophysical neutrino sources
from $p-\gamma$ interactions.}
\label{fig:CR}
\end{figure}

Figure~\ref{fig:CR}  shows the cosmic ray fluxes measured by the
Fly's Eye~\cite{fly}, AGASA~\cite{agasa}, and
Yakutsk~\cite{yakutsk} experiments. The smooth curve shown in
Fig.~\ref{fig:CR} was used by us~\cite{WnB99} in setting a
conservative  upper
bound on the high energy neutrino fluxes. 
We now explain why the upper bound is robust and conservative.

The smooth curve was computed
assuming that in the nearby universe ($z = 0$) the energy production rate 
is 
\begin{equation}
\left(E_{CR}^2{d\dot N_{CR}\over dE_{CR}}\right)_{z~=~ 0}~=~
10^{44}{\rm erg\ Mpc}^{-3}{\rm yr}^{-1}.
\label{eq:ECR}
\end{equation}

It is possible that the energy generation rate given locally by
Eq.~(\ref{eq:ECR}) increases with redshift. In particular, we have
explained in Ref.~\cite{WnB99} how cosmic rays observed at earth to
have energies in excess of $10^{18}$~eV must have originated at small
redshifts because of the large energy loss rate at these high
energies.  In order to establish a conservative upper limit, we
assumed that the local rate given in Eq.~(\ref{eq:ECR}) evolves with
redshift at the maximum rate observed for any astronomical population,
i.e., the evolutionary rate exhibited by the
quasars~\cite{QSO,QSO1,QSOh}.  We also included the adiabatic energy
loss due to the expansion of the universe.

Figure~\ref{fig:CR} shows that the smooth curve which we have used to
estimate the cosmic ray flux above $10^{18}$~eV is a
conservative (i.e., high) estimate of the observed rate.\footnote{ 
We note that Fig.~\ref{fig:CR} shows that 
 the highest energy point measured by the AGASA experiment 
could be interpreted  to 
suggest (with $\sim ~ 1\sigma$ significance) that the  cosmic ray 
generation rate at $E>10^{20}$~eV is twice the 
rate obtained from our smooth curve generated by Eq.~(\ref{eq:ECR}), implying
that the upper bound might  be underestimated by a factor of two at 
$\sim10^{19}$~eV. 
However, the higher rate of generation  is not 
observed by the Fly's Eye and Yakutsk experiments, and even 
if correct would imply
only a small correction to the upper bound at this energy. }

\begin{figure}[!t]
\centerline{\psfig{figure=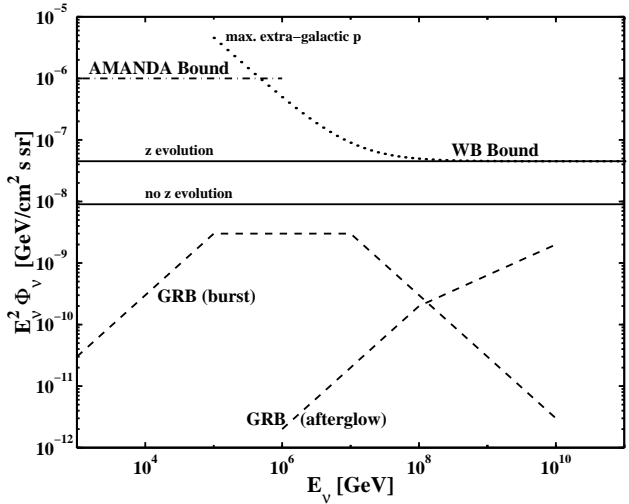,width=3.3in}}
\caption[]{The Waxman-Bahcall (WB) upper bound on muon neutrino
intensities ($\nu_\mu + \bar\nu_\mu$).  The numerical value of the
bound assumes that $100$\% of the energy of protons is lost to $\pi^+$
and $\pi^0$ and that the $\pi^+$ all decay to muons that also produce
neutrinos.  The WB upper bound exceeds the most likely neutrino flux
by a factor of $5/\tau$ for small optical depths $\tau$.  The upper
solid line gives the upper bound corrected for neutrino energy loss
due to redshift and for the maximum known redshift evolution (QSO
evolution, see text). In what follows, we will refer to this
conservative upper curve as the ``Waxman-Bahcall bound.''The lower
solid line is obtained assuming no evolution. The dotted curve is the
maximum contribution due to possible extra-galactic component of
lower-energy, $<10^{17}$~eV, protons as first discussed in
\cite{WnB99} (see Sec.~\ref{sec:lowenergy} for details). The
dash-dot curve shows  the experimental upper bound on diffuse
neutrino flux recently established by the AMANDA experiment
\cite{amanda-bound}. The dashed curves show the predictions of the GRB
fireball model \cite{WnB97,WnB99,WnB00}.  }
\label{fig:bound}
\end{figure}

What is the neutrino bound that results from the observed cosmic ray
flux?  Figure~\ref{fig:bound} shows the numerical limit that is
implied by the cosmic ray observations. The upper horizontal curve is
computed by assuming that the cosmic ray sources evolve as rapidly as
the most rapidly evolving known astronomical sources.  This very
conservative limit is what we shall mean when in the following we
refer to the `Waxman-Bahcall bound.'  The lower curve is computed
assuming that the number density of cosmic ray sources at large
distances is the same as in the local universe. We will discuss the
implications of the bound for previously published AGN models
in Sec.~\ref{sec:AGN}.

\section{Beyond $10^{20}$ eV}
\label{sec:nogzk}

The theoretical curve in Fig.~\ref{fig:CR} shows the predicted decrease 
above $5\times 10^{19}$ eV in the 
observed cosmic ray flux due to interactions with the cosmic microwave 
background~\cite{gzk}.  The existing obervations do not clearly 
demonstrate this decrease and therefore many authors have speculated that
there may be a new source of ultra-high energy neutrinos beyond $10^{20}$ eV. 
  
In our original discussion (see Fig.~1 of Ref.~\cite{WnB99} or 
Fig.~\ref{fig:bound} of the present paper), we did not 
extend the upper bound to energies beyond $10^{20}$ eV because 
we judged that 
too little is presently known about the cosmic ray flux in that region. 
However, as a useful guide to designing experiments,
it is a reasonable expectation that the production rate will not exceed 
by a large factor 
the flux predicted by an extension of the $E^{-2}$ production rate used in 
setting the limit at observed proton energies of $10^{19}$ eV.  
 
In our view, it is reasonable to extend the Waxman-Bahcall limit
beyond $10^{20}$ eV by simply extrapolating the horizontal line in
Fig.~\ref{fig:bound} to higher energies.  The validity of this
extrapolation will be tested by future measurements of the spectrum of
ultra-high energy cosmic rays, measurements that will determine
observationally whether the spectrum exhibits the predicted
Greisen-Zatsepin-Kuzmin (GZK) cutoff at high energies.

We have suggested elsewhere~\cite{BWn00} that the clustering of
sources of ultra-high energy ($> 10^{20}$) cosmic rays may give rise
to more events beyond the GZK cutoff than expected on the basis of a
homogeneous distribution of sources. Other authors (see, for example,
the references cited in Sec.~\ref{sec:exceed}) have suggested
exotic sources of ultra-high energy cosmic rays that are not described
by the $E^{-2}$ proton generation rate.  If any of these scenarios is
correct, then the upper bound shown in Fig.~\ref{fig:bound} will
have to be raised above the simple extrapolation we currently advocate
as a useful guide. However, the existing cosmic ray data suggest that
the factor by which the WB upper bound is conservative, $5/\tau$, will
more than compensate for any increase in cosmic ray flux beyond the
GZK limit.

\section{Lower energy: $E<10^{18}$~eV}
\label{sec:lowenergy}

Figure~\ref{fig:CR} 
shows that our smooth curve for the
extragalactic 
proton cosmic ray flux  falls
below the total observed flux  for  energies less than $10^{19}$~eV.
We have assumed (see
Eq.~\ref{eq:ECR}) that the generation rate of extragalactic protons
is proportional to  $E^{-2}$, where $E$ is the proton  energy.
This $E^{-2}$ dependence  is produced generically by the Fermi mechanism 
for accelerating high energy cosmic rays in shocks~\cite{accelerate}. 
If the lower energy cosmic rays were protons from
extragalactic sources, then (as we noted earlier~\cite{WnB99}) one
could raise the upper bound for astrophysical neutrinos at these 
lower energies.

Could the lower energy cosmic rays be primarily extragalactic protons?
Unfortunately, the answer is no\footnote{In the original version of
their paper, astro-ph/9812398 v1, Mannheim, Protheroe, and Rachen
assumed that $100$\% of the lower energy cosmic rays could be
extragalactic protons. In the third version of their paper, v3, they
accepted the argument presented in the first version of our paper
hep-ph/9902383 v1, and repeated here, that only a small fraction
of these lower energy cosmic rays are extragalactic protons.}. The
available observational evidence suggests that only a small fraction of the
cosmic-ray flux in the energy range of $10^{14}$~eV to $10^{17}$~eV is
composed of protons.  Direct (balloon) composition measurements at
$10^{14}$~eV \cite{JACEE} show that the fraction of cosmic-ray flux
composed of protons at this energy is $\sim20\%$.  Air-shower and
cosmic-ray tracking detectors measurements indicate~\cite{HEGRA} that
the proton fraction decreases in the energy range of $10^{14}$~eV to
$10^{16}$~eV. In fact, the Fly's Eye and AGASA experiments
support~\cite{Dawson98} a composition strongly dominated by
(consistent with 100\%) heavy nuclei at $10^{17}$~eV.  Moreover, both
the Fly's Eye~\cite{Bird98} and the AGASA~\cite{Hayashida98}
collaborations have reported, for energies less than $3 \times
10^{18}$~eV, a small but statistically significant enhancement of the
cosmic ray flux near the Galactic plane.  This enhancement is expected
for a Galactic, but not an extragalactic, origin for the cosmic rays
in this energy domain.

In summary, the observational evidence is that most of the observed
cosmic rays in the energy range $10^{14}$~eV to $10^{17}$~eV are not
protons and therefore the total flux of cosmic rays cannot be used to
raise the upper bound for neutrinos in the energy range $10^{14}$~eV
to $10^{17}$~eV. The NASA satellite ACCESS is expected to
provide accurate measurements of the composition of the cosmic rays at
energies up to $\sim 10^{15}$ eV sometime in the next decade.

Assuming, conservatively, that $\sim 10$\% of the
cosmic rays in this energy region are protons, the neutrino bound may
be raised at energies $E_\nu<10^{16}$~eV (Since the cosmological model
we have used, see Eq.~\ref{eq:ECR} and Fig.~\ref{fig:CR}, accounts for
$\sim 10$ \% of the total cosmic ray flux at $10^{17}$~eV, and a
progressively larger fraction at higher energies, one cannot increase
the upper bound on neutrino fluxes at energies $\geq 10^{16}$~eV).
Since the energy density in cosmic rays is approximately proportional
to $E^{-1}$ for energies less than $10^{18}$~eV, the actual
extragalactic neutrino flux could exceed the most stringent
Waxman-Bahcall limit by a factor of $(10^{16}$~eV/$E_\nu$) for
neutrino energies $E_\nu < 10^{16}$~eV.

Figure~\ref{fig:bound} illustrates 
with a curved dotted line 
the maximum contribution from unrecognized extragalactic sources of 
protons, i.e., sources of extragalactic protons that do not contribute 
significantly to the observed cosmic ray flux at $10^{19}$ eV.
The `unknown source' contribution is obtained by multipling the proton 
generation rate that is proportional to $E^{-2}$ by an analytic 
approximation to the  
maximum additional contribution allowed at lower energies by the existing 
experiments.  
Thus
\begin{equation}
{\rm max. ~ proton ~rate} \propto E^{-2}
\left[ 1 ~+~ 0.1((10^{17} eV)/E) \right],
\label{eq:maxprate}
\end{equation}
where $E$ is the proton energy. The recent experimental bound established by
the AMANDA experiment already rules out a contribution  
from unrecognized extragalactic sources of protons at the maximum level 
allowed by present cosmic-ray observations.

We note that including a contribution to the neutrino flux due to
neutrino production by  photo-meson interactions on heavy 
nuclei will not affect the neutrino bound. For heavy elements,
the cross section for photo-dissociation is higher than, 
and its energy threshold is lower than,
for photo-meson production. Thus, heavy elements will dissociate before
losing a significant fraction of their energy to neutrino production.

\section{AGN models}
\label{sec:AGN}

What predictions have been made for the neutrino fluxes from AGN jets? 
Are the fluxes large enough to be measurable in a practical detector?

There are two classes of models that have been considered as
explanations for 
the emission of radiation from AGN
jets, ones in which radiation is due to electromagnetic
processes~\cite{em} and 
others in which emission of radiation
involves the acceleration of very high energy
protons~\cite{Mannheim95,Protheroe96,Halzen97}.  
For the first class of models, no significant flux of high energy
neutrinos is expected. The second class of models has 
been normalized by postulating that the jets produce the observed
gamma-ray background via photo-meson processes on high energy protons,
which produce $\pi^0$'s that decay into gamma rays. The photo-meson
process, if it produces the gamma-ray background, would
also  produce a large flux of high energy neutrinos via charged
meson decay. Therefore, the second class of models, involving
photo-meson interactions on high energy protons, has 
received a lot of attention, especially from particle experimentalists.

The previously published AGN models that are  discussed in the following
subsection have been used to estimate the expected event rates in the
AMANDA, ANTARES, Auger, ICECUBE, NESTOR, and OWL
detectors~\cite{antares,nestor,baikal,amanda,owl,auger,watson}.
However, all of these models exceed by more than an order of magnitude
the maximum neutrino flux permitted by the WB bound (see
Fig.~\ref{fig:agnmodels}).

\subsection{Previously published models}
\label{subsec:previous}

\begin{figure}[!b]
\centerline{\psfig{figure=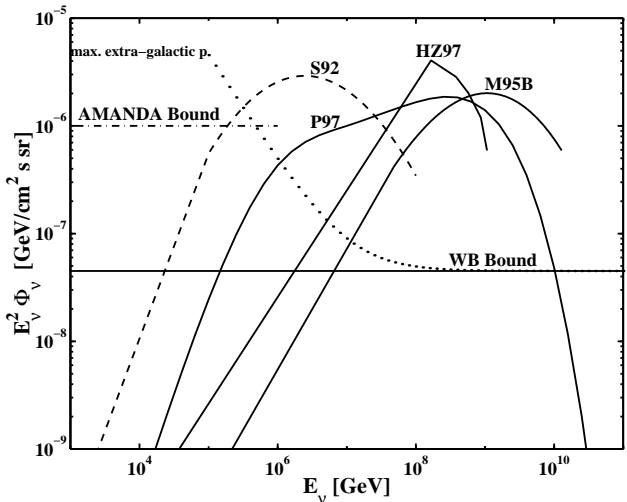,width=3.3in}}
\caption[]{The Waxman-Bahcall (WB) upper bound on muon neutrino
intensities ($\nu_\mu + \bar\nu_\mu$)
(solid line) compared to predictions of representative 
AGN jet models,
taken from the earlier papers of Mannheim~\cite{Mannheim95}
(marked M95B in the figure), 
Protheroe~\cite{Protheroe96} (P97), and  
Halzen and Zas~\cite{Halzen97} (HZ97). 
The AGN models were normalized so that the calculated gamma-ray flux 
from $\pi^0$ decay  fits the observed gamma-ray background. 
The AGN hidden-core conjecture (S92), to which the WB upper bound does not 
apply due to high photo-meson optical depth of the source, 
is taken from \cite{Stecker}. Note, that this conjectrure is already ruled
out by the AMANDA upper bound \cite{amanda-bound}.}
\label{fig:agnmodels}
\end{figure}

Figure~\ref{fig:agnmodels}  compares the Waxman-Bahcall 
 upper bound to predictions by various proton acceleration models for
 AGN jets.  The previously published models illustrated in 
Fig.~\ref{fig:agnmodels} are inconsistent with the cosmic ray limit; they 
typically predict fluxes one to two orders of magnitude above the 
Waxman-Bahcall upper bound.

The expected detection rate in a km$^2$ detector (e.g., ANTARES,
ICECUBE, or NESTOR) is less than one event per year from AGNs with
spectral neutrino energy shapes like the models P97, HZ97, and M95B shown in
Fig.~\ref{fig:agnmodels} if
a neutrino flux consistent with the Waxman-Bahcall bound is assumed.
This rather pessimistic result can be obtained from Tables III--VI of
Ref.~\cite{Gandhi98} by dividing by $30$ the detection rate given in
those tables for the model denoted by AGN-M95. The same model is
labeled `M95B' in Fig.~\ref{fig:agnmodels}, where one can see that the
flux predicted from M95B is about a factor of $30$ above the cosmic
ray upper bound.

The neutrino event rate in the Auger detector`\cite{auger,watson} is
 less than or of order $1$ event per year above a neutrino energy of
 $10^{19}$ eV for optically thin sources, i.e., sources that satisfy
 the WB bound. The rate may be two orders of magnitude larger if the
 OWL/AIRWATCH detector is built according to the preliminary
 specifications~\cite{owl}.

Since $p-\gamma$ reactions give rise to  high energy gamma rays (via
$\pi^0$ decay) as well as to high energy neutrinos (via charged pion
decay), the flux of neutrinos is proportional to the flux of gamma rays. 
Since the models with high neutrino fluxes that are 
shown in Fig.~\ref{fig:agnmodels} were all
chosen by their authors so that they could account for the observed
gamma ray flux, it is clear that the still allowed AGN models 
(with the $30$ times lower 
neutrino flux) can only account for a few percent   of the gamma ray
background. 

\subsection{Are AGNs optically thick to photo-meson
interactions?}
\label{sec:thick}

The WB bound applies only to sources which are optically thin to
photo-meson interactions\footnote{The photo-meson optical depth of high-energy
protons in  GRBs is discussed in~\cite{WnB99,WnB97}.
It is shown there that the observation of high energy, $\sim100$~MeV, 
photons in GRBs implies that the sources are not optically thick to 
photo-meson interactions.  In the context of Fireball models for GRBs,
the optical depth to nucleon-meson interactions is also
 small~\cite{WnB97,BM2000}.}.  As mentioned above, previously published
AGN jet models satisfy this requirement. One may argue~\cite{MPR99},
however, that new models may be constructed to have large optical
depth and nevertheless fit the available data.

We discuss 
this possibility below. We first explain why published AGN
models predict small optical depth, and then discuss whether new, high
optical depth models can be constracted.

The optical depth to photo-messon interactions for protons of energy 
$\epsilon_p$ can be related to the optical depth for pair production of 
photons of much lower energy, $\epsilon_\gamma$.  The threshold relation 
for photon-meson interactions is $\epsilon_p \, \epsilon_\gamma \approx
m_\pi \, m_p$ and the threshold relation for pair production is
$\epsilon_\gamma^\prime \, \epsilon_\gamma =2 m_e^2$.  Thus,
as first pointed out in \cite{WnB99}, a photon of energy 
$\epsilon_\gamma$ that causes a photon-meson interaction with a proton of 
energy $\epsilon_p$ could also  pair-produce with a photon of energy 
$[2m_e^2/(m_\pi m_p)]\epsilon_p = 4 \times 10^{-6} \epsilon_p $.
Taking account of the ratio between photo-meson and pair production cross 
sections, one finds that 
\begin{equation}
\tau_{\gamma \, p} ( \epsilon_p ) ~\approx 5\times10^{-4}~ 
\tau_{\gamma \, \gamma}(\epsilon_\gamma = 4\times 10^{-6} \epsilon_p).
\label{eq:tauphotovspair}
\end{equation}  
Observed AGN photon energy distributions typically follow a power-law,
$dn_\gamma/d\epsilon_\gamma\propto \epsilon_\gamma^{-2}$. For such 
photon spectra, one can easily show~\cite{WnB99} 
that $\tau_{\gamma \, \gamma} \propto \epsilon_\gamma$, and hence that
\begin{equation}
\tau_{\gamma \, p} ( \epsilon_p ) ~\approx 2 
\tau_{\gamma \, \gamma}(\epsilon_\gamma = 10{\rm GeV})
\left({\epsilon_p\over10^{19}{\rm eV}}\right).
\label{eq:tau}
\end{equation}

Emission of $\sim1$~TeV 
photons from ``blazars,'' AGN jets nearly aligned with our line of sight,
is now well established \cite{eTeV},
and there is evidence that the high-energy photon spectrum extends as a 
power-law at least to $\sim10$~TeV \cite{TeV}. The high energy photon 
emission is the main argument used \cite{pionAGN} in 
support of the hypothesis that high-energy emission from blazars
is due to pion decay rather than inverse Compton scattering. However,
the observation of $>1$~TeV photons implies, as shown by Eq. 
(\ref{eq:tauphotovspair}), that the jet optical depth to photo-meson
interaction is very small. Indeed, one can readily see from Fig.~4 of 
Ref.~\cite{Mannheim96} that  high-energy photon data
of northern hemisphere blazars requires small pair-production
optical depth at photon energy $<0.1$~TeV. All protons blazar
models shown in the figure have $\tau_{\gamma \, \gamma}<1$  at this energy,
i.e. $\tau_{\gamma \, p}<1$ for $\epsilon_p <  10^{19}$ eV.

TeV emission is observed from the nearest blazars, which are hence
relatively low-luminosity blazars.  
It had recently been argued by Mannheim, Protheore and Rachen
(MPR, \cite{MPR99}), that the emission of
high energy, $\sim1$~TeV, photons may be suppressed in high luminosity 
blazars, for which models with high optical depth may therefore be 
constructed.
The most intense high-energy gamma-ray source is 3C 279.  Figure~6 of 
Ref.~\cite{Hartman96}, which is reproduced here as Fig.~\ref{fig:3c279},
shows that the measured optical depth for pair production is at most 
$\tau_{\gamma \, \gamma} \sim 1$ at $\epsilon_\gamma = 10$ GeV. Combining 
this observation with Eq.~(\ref{eq:tau}) we see that 
$\tau_{\gamma \, p} ( \epsilon_p ) < 1$ for $\epsilon_p <  10^{19}$ eV.
Making the {\it ad hoc} assumption, that the pair-production optical depth 
for high luminosity blazars like  3C~279 exceeds unity 
for photons above 10~GeV, i.e. just above the highest energy for which 
data are available (see Fig.~\ref{fig:3c279}),
MPR concluded that the
AGN neutrino flux may slightly exceed the WB bound at $\sim10^{18}$~eV
(see dotted curve in their Fig.~5a).

\vbox{
\begin{figure}[!bh]
\centerline{\psfig{figure=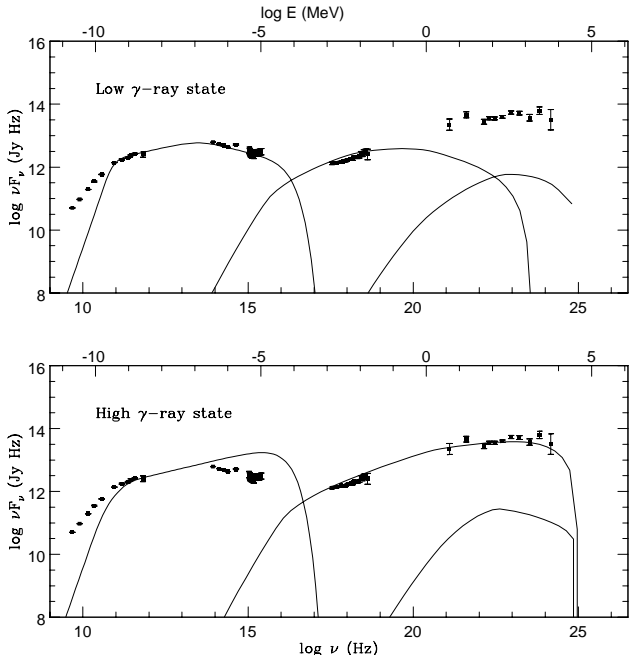,width=3.5in}}
\vglue-.2in
\caption[]{The emission spectrum of 3C~279 over 15 decades in
frequency, adatped from Ref.~\cite{Hartman96}.  The measured fluxes
are plotted as small squares in the figure. The flare in June 1991 was
studied by many different groups observing in a large number of
wavelength domains. Out to the last measured point at about $10$ GeV,
the emission spectrum shows no evidence for turning over due to a
large optical depth.  The solid curves are the predictions for
emission from a uniform relativistic moving sphere; more complex
models are discussed in ~\cite{Hartman96}.}
\label{fig:3c279}
\end{figure}
}

In order to avoid the constraint given in Eq. (\ref{eq:tau}), 
and construct AGN models with significant 
optical depth to photo-meson interaction, 
one must argue that the spectral distribution of the background photons 
(for photo-meson and pair-production intercations) deviates from a
$dn_\gamma/d\epsilon_\gamma\propto \epsilon_\gamma^{-2}$ power-law. Since the
observed spectral distribution does follow an $\epsilon_\gamma^{-2}$ 
power-law, one must argue that the background radiation with 
which high energy photons and protons may interact is not the radiation
associated with the observed photon flux. This may indeed be the case for
jets which expand with high Lorenz factor $\Gamma$. In this case, an 
ambinet isotropic radiation field, $u_{amb.}$, through which the jet
expands, will dominate pair-production and photo-meson interactions 
as long as its energy density exceedss a fraction $\Gamma^{-4}$
of the energy density associated with the observed radiation,
$u_{obs.}$, which is produced within the jet. The presence of such radiation
field can not be ruled out. However, the energy density of such ambinet
radiation is limited to values not much exceeding 
$u_{amb.}/u_{obs.}=\Gamma^{-4}$, since for $u_{amb.}/u_{obs.}>\Gamma^{-4}$ 
inverse Compton scattering emission of jet electrons would dominate over
the synchrotron emission which produces the observed radiation in the 
radio to UV bands.

In order to obtain the maximum possible optical depth, MPR have
therefore adopted \cite{MPR99} the following {\it ad hoc} assumptions:
(1) The pair production optical depth in luminous blazars, e.g. 3C~279,
exceeds unity for photons above 10~GeV; (2) The optical depth is due
to ambient radiation (which is not detected); (3) The spectrum of the
ambinet radiation deviates from a $\epsilon_\gamma^{-2}$ power law in
such a way that its energy density increases by an order of magnitude
at photon energies just below those corresponding to the
pair-production threshold of 10~GeV photons. Adopting these
implausible assumptions, MPR concluded that the AGN neutrino flux may
exceed the WB bound by a factor $\sim2$ at $\sim10^{18}$~eV (see thick
dahsed curve in their Fig.~5a).

\section{Can one evade the upper bound with the aid of magnetic
fields?}
\label{sec:magnetic}

In Sec.~III of 
Ref.~\cite{WnB99} we considered various ways that one might try to
avoid the upper limit on the flux of high energy neutrinos by invoking
magnetic fields. In particular, we discussed the possibility that
the cosmic-ray density observed at Earth is lower than the average
universe cosmic-ray density, due to confinement of protons by magentic
fields surrounding the cosmic-ray sources, or due to large-scale structure
fields. We have shown that, given observational constraints on such magentic
fields, magnetic proton confinement can not affect the high-energy
cosmic-ray flux, and therefore can not affect the neutrino bound.

In Sec.~IIIc of Ref.~\cite{WnB99} we have discussed in detail the
possible effects of large-scale structure magnetic fields, and demonstrated
that even if such fields are built to equipartition levels, protons of energy
exceeding $\sim10^{16.5}$~eV can not be confined to large-scale structures 
over a Hubble time. Large-scale structure magnetic fields can not therefore
affect the WB bound at energy $\gtrsim10^{15}$~eV.
It should be pointed out here, that had proton confinement by large scale
structure magnetic fields been possible, which might be the case for 
low energy protons, it would have most likely implied
that the neutrino upper bound should be lower, rather than higher
at $\epsilon_\nu\lesssim10^{15}$~eV. This 
is due to the fact that confining high-energy cosmic-rays to high density
regions would imply that the local cosmic-ray density is higher than the
universe average. Since, on the other hand, neutrinos can not be confined
and their density is uniform, their flux would be bound to a lower value
than implied by our assumption of a uniform cosmic-ray density.

Recently, it has been argued by Mannheim, Protheore and Rachen (MPR,
\cite{MPR99}), that while high energy protons can not be confined to
the vicinity of their sources by magnetic fields, adiabatic losses of
protons escaping magnetic halos surrounding AGN jets may lead to
significant decrease in proton energy for $\epsilon_p<10^{18}$~eV,
which may lead to modification of the bound for
$\epsilon_\nu<10^{17}$~eV. Clearly, even if this claim is correct,
proton AGN models are still in contradiction with the Waxman-Bahcall
bound, since the bound is not affected at $\epsilon_\nu>10^{17}$~eV
(see Fig.~\ref{fig:agnmodels}).  Moreover, the key arguments made by
MPR in deriving the claimed adiabatic energy loss are flawed.  First,
the magnetic field model adpoted by MPR is in conflict  with
observations.  While the jet magnetic field inferred from observtaions
is high, tens of $\mu$G (e.g. \cite{Daly95}), depolarization
measurements \cite{Garrington91} imply small halo magnetic field,
0.1--1~$\mu$G at the central halo region, $r\sim10$~kpc, and much
smaller at large distances, $r\sim100$~kpc. MPR assume a spherically
symmetric field structure with amplitude decreasing as $r^{-2}$ from
30~$\mu$G $r=10$~kpc to 0.3~$\mu$G at $r=100$~kpc. Second, MPR assume
that high energy protons are produced in the central region of the
jet, where the magnetic field is high, and then are confined to jet
plasma which expands adiabatically to the halo plasma conditions of low
magnetic field. This is an implausible  description of
the plasma evolution and  of proton confinement.  Acceleration
of protons to high energy is typically expected to occur at the outer
edge of the jet at the strong shock.  Moreover, since the jet is
narrow, typical opening angles are 10 degrees, the typical jet
structures are $\sim1$~kpc in scale, and neutrons produced by
photo-meson interactions leading to $\nu$ productiuon escape the jet
before decaying.

\section{Contributions from non-resonant interactions}
\label{sec:nonresonant}

The high energy-neutrino flux from GRBs was calculated by Waxman \&
Bahcall \cite{WnB97,WnB99} using the ``$\Delta$-approximation,''
i.e. assuming that photo-meson production is dominated by interaction
of protons with photons of energy close to that corresponding to the
$\Delta$-resonance.  It has recently been argued by M\"ucke {\it et al.}
\cite{Mucke98} that for characteristic GRB photon spectrum, there is
an additional contribution to photo-meson production from interaction
of protons with photons of energy much higher than that corresponding
to the resonance, leading to significant deviation from results
derived based on the $\Delta$-approximation.  We point out here, that
the contribution to neutrino production of non-resonant interactions
is small, leading to only negligible modification of results obtained
using the $\Delta$-approximation.

The GRB photon spectrum is well fitted in the
BATSE detector range, 30~keV--3~MeV, by a combination
of two power-laws, 
$dn_\gamma/d\epsilon_\gamma\propto\epsilon_\gamma^{-\beta}$ 
with different values of $\beta$ at low and high energy \cite{Band}. The
break energy (where $\beta$ changes) in the observer frame is typically 
$\epsilon^{\rm ob.}_{\gamma b}\sim1{\rm MeV}$, 
with $\beta\simeq1$ at energies below the break and $\beta\simeq2$ 
above the break (the plasma emitting the GRB radiation expands with 
Lorentz factor $\Gamma\simeq300$, and particle energies in the plasma
frame are smaller by a factor $\Gamma^{-1}$ compared to observed particle
energies). For protons of observed energy 
$\epsilon_p\le10^{16}$~eV, the (observed) photon threshold energy for 
photo-meson interaction is higher than the break energy 
$\epsilon^{\rm ob.}_{\gamma b}\sim1{\rm MeV}$ \cite{WnB97}. Such protons
therefore interact only with the steep part of the photon spectrum,
$dn_\gamma/d\epsilon_\gamma\propto\epsilon_\gamma^{-2}$.
Thus, for $\epsilon_p\le10^{16}$~eV 
the contribution from interaction with high
energy photons, i.e. photons of energy well above the threshold and therefore
well above the energy corresponding to the $\Delta$-resonance, 
is negligible simply
because there are very few photons of such high energy. 

Protons of energy $\epsilon_p\gg10^{16}$~eV
may interact with photons well below the break,
where the photon spectrum is flatter, 
$dn_\gamma/d\epsilon_\gamma\propto\epsilon_\gamma^{-1}$.
In this case there may
be significant contribution to photo-meson production from interaction 
of protons with photons of energy well above the threshold, and therefore
well above the energy corresponding to the $\Delta$-resonance, since the
photon spectrum is flat over significant energy range, from the threshold
energy to $\epsilon^{\rm ob.}_{\gamma b}\sim1{\rm MeV}$.
However, contribution from non-resonant interaction becomes 
significant only at very high energy: for protons of energy 
$\epsilon_p>10^{18}$~eV photo-meson production of pions is increased by 
only a factor of $\sim2$ compared to the production rate based on the 
$\Delta$-approximation. 
Moreover, this increase has no effect on the neutrino production discussed by 
Waxman \& Bahcall \cite{WnB97,WnB99}, since the high energy pions 
produced in interaction of protons with $\epsilon_p>10^{18}$~eV lose their
energy through synchrotron emission before decaying, and therefore
do not contribute to the neutrino flux, which is strongly suppressed
at observed neutrino energy $>10^{16}$~eV
\cite{WnB97,RnM98,WnB99}.

\section{Discussion}
\label{sec:discussion}

In this paper, we have shown that the Waxman-Bahcall upper limit on
neutrino fluxes is robust for neutrino energies above $10^{16}$~eV.
Cosmic ray experiments set a firm upper limit on the flux of
extragalactic protons, and hence on the flux of extragalactic
neutrinos, that are produced by  $p + \gamma$ or $p-p$ interactions in
astronomical sources that are optically thin to photo-meson and
nucleon-meson interactions on high energy protons.  
The upper solid curve in Fig.~\ref{fig:bound}, which is labeled WB
bound, is a conservative representation of the limit imposed by the
cosmic ray observations and includes the maximum plausible 
redshift evolution
of the sources of high-energy cosmic rays.  The WB bound exceeds the
most likely neutrino flux from $p + \gamma$ or $p-p$ interactions by a
factor of $5/\tau$, for small optical depths, $\tau$, to photo-meson
and nucleon-meson interactions.

The Waxman-Bahcall upper bound is about two orders of magnitude less
than published predictions~\cite{Mannheim95,Protheroe96,Halzen97} of
neutrino fluxes expected from models for AGN jets that explain the
gamma-ray background by photo-meson interactions on high energy
protons (see Fig.~\ref{fig:agnmodels}).  Given the neutrino upper
limit, photo-mesonic interactions in optically thin sources like AGN
jets can at most account for a few percent of the flux of the
gamma-ray background~\cite{WnB99}.

The Waxman-Bahcall bound implies an upper limit to the expected event
rate from optically thin sources that may be observed in the
AMANDA, ANTARES, Auger, ICECUBE, NESTOR, and OWL/AIRWATCH
detectors~\cite{antares,nestor,baikal,amanda,owl,auger,watson}.  In
${\rm km^2}$ detectors like ANTARES, ICECUBE, or NESTOR, one expects
to detect less than or of order $1$ neutrino event per year 
from optically thin AGNs with neutrino spectral
energy shapes like the AGN models P97, HZ97, and M95B shown in
Fig.~\ref{fig:agnmodels}.  For the Auger detector, one expects less
than or of order $1$ neutrino event per year above $10^{19}$ eV from
all sources satisfying the WB bound. The event rate could be $\sim$
two orders of magnitude larger for the OWL detector if it performs
according to the preliminary specifications.  These estimates are
upper limits and do not include the factor of $5/\tau$ by which, for
small optical depths $\tau$, the WB bound exceeds the plausible
expected neutrino flux.  The rate estimates given in this paragraph
are much less than the estimates given in the published papers
describing the capabilities of the neutrinos detectors, since in those
papers AGN models like those shown in Fig.~\ref{fig:agnmodels} were
assumed in the calculations.

At neutrino energies below $10^{16}$
eV, the flux of extra-galactic neutrinos from optically thin sources
may reach as high as the dotted curve in Fig.~\ref{fig:bound}. In
order for this curve, labeled `max. extra-galactic p' in
Fig.~\ref{fig:bound}. to apply, a large fraction of the cosmic rays
with energies below $10^{17}$ eV must be protons of extra-galactic
origin (cf. Eq.~\ref{eq:maxprate}). For cosmic rays more energetic
than $10^{20}$ eV, there is not yet an accurate
determination of the cosmic ray flux.  Hence there is no way of
rigorously extending the WB bound for neutrino energies $> 10^{19}$
eV. We propose that horizontal line in Fig.~\ref{fig:bound} simply be
extrapolated to higher energies as a plausible bound if the source
spectrum continues to fall off at ultrahigh-energies  as $E^{-2}$. 

The large area neutrino
experiments~\cite{antares,nestor,baikal,amanda} that are currenty
being constructed are designed to detect neutrinos with energies $<
10^{15}$ eV, corresponding to protons with energies less than or of
the order of $10^{16}$ eV. The WB limit is constructed by normalizing
to the observed flux of cosmic ray protons at $10^{19}$ eV and
extrapolating to lower energies using the $E^{-2}$ source
spectrum. The extrapolation is plausible but not rigorous for the
neutrino energy domain of the ANTARES, BAIKAL, ICECUBE, and NESTOR
experiments.

There are two hypothetical ways of exceeding the Waxman-Bahcall limit
without violating the constraint imposed by cosmic ray
observations.  First, the neutrinos could be produced in sources that
are optically thick to photo-nucleon or nucleon-nucleon interactions.
Second, the neutrinos could be produced by processes that do not give
rise to high energy cosmic rays, such as the decay of dark matter
particles, topological defects, superheavy relic neutrinos, or
ultrahigh-energy photons.  So far, there is no obervational evidence
supporting any of these possiblities.

We have investigated a number of {\it ad hoc} scenarios invented to try
to find ways of violating the Waxman-Bahcall bound. None of the
suggested scenarios, including various ideas involving magnetic
fields, provide a physically self-consistent mechamism for raising the
upper limit to the neutrino flux implied by a straightforward
interpretation of the cosmic ray observations.

\acknowledgments
We are grateful to Dr. R. C. Hartman for providing us with a
postscript file of Fig.~\ref{fig:3c279}. JNB acknowledges the
hospitality of the Weizmann Institute during a visit where much of
this work was done and also partial
support from NSF grant 
No. PHY-0070928 to the Institute for Advanced Study.  EW acknowledges
the hospitality of the Institute for Advanced Study during a visit where
this work was completed, and 
support from BSF grant 98-00343 and AEC grant 38/99.

\end{document}